\documentclass[10pt,aps,prl,twocolumn,preprintnumbers,amsmath,amssymb,floatfix,showpacs,citeautoscript,superscriptaddress]{revtex4-2}
\usepackage[latin1]{inputenc}
\usepackage[T1]{fontenc}
\usepackage[english]{babel}
\usepackage[pdftex]{graphicx}
\usepackage{amsmath}
\usepackage{times}
\usepackage{physics}
\usepackage{color}
\usepackage[usenames,dvipsnames,svgnames,table]{xcolor}
\usepackage[colorlinks,plainpages=false,linkcolor=blue,urlcolor=blue,citecolor=blue,pdfpagemode=UseNone]{hyperref}
\usepackage{svg}
\usepackage[normalem]{ulem}

\renewcommand{\AA}{\text{\r{A}}}

\newcommand{\mbf}[1]{\mathbf{#1}}

\newcommand{\bk}{\mbf{k}}
\newcommand{\bq}{\mbf{q}}
\newcommand{\bR}{\mbf{R}}
\newcommand{\bRp}{{\mbf{R}}^{\prime}}
\newcommand{\br}{\mbf{r}}

\newcommand{\BSCCO}{{Bi$_2$Sr$_2$CaCu$_2$O$_{8+\delta}$}}

\definecolor{d12orange}{rgb}{0.8500    0.3250    0.0980}
\definecolor{g8yellow}{rgb}{0.    0.6    0.298}
\definecolor{ppurple}{rgb}{0.4940    0.1840    0.5560}

\definecolor{mag}{RGB}{255,0,255}

\begin{document}

\title
{
\boldmath
{Theory of Josephson scanning microscopy with $s$-wave tip on unconventional superconducting surface: application to Bi$_2$Sr$_2$CaCu$_2$O$_{8+\delta}$ }
}

\author{Peayush Choubey}
\affiliation{Department of Physics, Indian Institute of Technology Roorkee, Roorkee 247667, Uttarakhand, India}

\author{P. J. Hirschfeld}
\email{pjh@phys.ufl.edu}
\affiliation{Department of Physics, University of Florida, Gainesville, Florida 32611, USA}

\date{\today}

\begin{abstract}
Josephson scanning tunneling microscopy (JSTM) is a powerful probe of  the local superconducting order parameter, but studies have been largely limited to cases where superconducting sample and superconducting tip both have the same gap symmetry- either s-wave or d-wave.   It has been generally assumed that in an ideal $s$-to-$d$ JSTM experiment the critical current would vanish everywhere, as  expected for ideal $c$-axis planar junctions. We show here that this is not the case.  Employing first-principles Wannier functions for Bi$_2$Sr$_2$CaCu$_2$O$_{8+\delta}$, we develop a scheme to compute Josephson critical current ($I_{c}$) and quasiparticle tunneling current measured by JSTM with sub-angstrom resolution. We demonstrate that the critical current for tunneling between an s-wave tip and a superconducting cuprate sample has largest magnitude above O sites and it vanishes above Cu sites. $I_{c}$ changes sign under $\pi/2$-rotation and its average over a unit cell vanishes, as a direct consequence of the $d$-wave gap symmetry in cuprates. Further, we show that $I_{c}$ is strongly suppressed in the close vicinity of a Zn-like impurity owing to suppression of the superconducting order parameter. More interestingly, $I_{c}$ acquires non-vanishing values above the Cu sites near the impurity.The critical current modulations produced by the impurity occur at characteristic  wavevectors distinct from the quasiparticle interference (QPI) analogue. Furthermore, the quasiparticle tunneling spectra in the JSTM set-up shows coherence peaks and impurity-induced resonances shifted by the $s$-wave tip gap. We discuss similarities and differences in JSTM observables and conventional STM observables, making specific predictions that can be tested in future JSTM experiments.
\end{abstract}

\maketitle

\section{Introduction}
\label{sec:Intro}
The electron pair amplitude in a superconductor is the ultimate quantum mechanical fingerprint for the many body state. Its symmetry and structure give clues to the physical origin of the processes that bind the pair together. One great goal of superconductivity research is to probe the nature of the superconducting state by imaging this pair condensate directly.

Recent technical advances have resulted in dramatic improvement in JSTM, allowing for  high-resolution maps of the critical current. The technique has recently been applied to the underdoped Bi$_2$Sr$_2$CaCu$_2$O$_{8+\delta}$ (BSCCO) to study {a putative} pair density wave (PDW) \cite{hamidian2016detection,Chen2022} and the interplay of electron pair density and charge transfer energy scale \cite{Mahony2022}. The technique has provided information on the atomic scale variation of the superfluid condensate in Pb \cite{Yazdani2016}  and the iron-based superconductor FeTe$_{0.55}$Se$_{0.45}$ (FeTeSe) \cite{cho2019strongly}. Ref. \cite{Yazdani2016} studied the critical current $I_c$ near magnetic impurities, while the $I_c$ maps in Refs. \cite{hamidian2016detection,cho2019strongly} were shown to reflect strongly inhomogenous ``intrinsic" superconducting order. More recently, JSTM studies helped to identify possible PDW states in  on 2H-NbSe$_{2}  $\cite{Liu2021A,Liu2021B}, CsV$_{3}$Sb$_{5}$ \cite{Chen2021} and UTe$_{2}  $\cite{Gu2023}.

The Pb and Fe(Se,Te) measurements used Pb tips, such that the tunneling in both these cases was expected to be between an $s$-wave tip and sample surface. The simplest theory of JSTM \cite{smakov2001josephson} assumed a pointlike tip and constant matrix elements for tunneling. In these experiments, one should therefore observe a nonzero critical current proportional to the product of the two $s$ wave order parameters of tip and sample, $\Delta_t\Delta_s$.  As in any SIS tunneling process, peaks from the quasiparticle branch are expected at a bias voltage $eV$ corresponding to $\Delta_t$ + $\Delta_s$. The inhomogeneous critical current in such a system in the presence of impurities, applicable in principle to Ref. \cite{Yazdani2016}, was discussed theoretically in Ref. \cite{graham2017imaging}.

By contrast, the BSCCO JSTM experiment \cite{hamidian2016detection} used an exfoliated BSCCO flake of roughly 10 \AA~ in size, such that the tunneling is  taking place between two $d$-wave superconductors, identical except for the finite size of the tip. In the limit of a pointlike tip, the critical current vanishes in the framework of Ref. \cite{smakov2001josephson}, essentially because at one spatial point the system cannot accommodate an intrinsically extended $d$-wave pair function.  However if one accounts for the nonzero size of the tip, as in Ref. \cite{graham2019josephson}, a nonzero critical current is expected. In real space, this occurs simply because of the tunneling between adjacent tip and sample bond pair amplitudes that have an identical phase structure.   

Josephson tunneling between a pure $s$-wave tip made of conventional superconductor and a $d$-wave cuprate (or other unconventional) superconducting sample could potentially provide a much simpler probe of the unconventional pair wave function. However, the symmetry mismatch dictates a vanishing critical current in both the momentum space-based theory of Ref. \cite{smakov2001josephson} and the real-space formalism\cite{graham2019josephson}. While JSTM of this type seems {\it a priori} useless, it is interesting to remember that over the years there have been many reports of planar Josephson tunneling between $s$-wave superconductors and cuprates. These have been explained by appealing to details of the tunneling process: surface roughness, or pits, or impurities, which destroy the momentum conservation of the pair at the interface \cite{Sun1994}. In addition, a JSTM experiment with a Pb tip on a BSCCO surface was discussed in Ref. \cite{Kimura2008}, and a Josephson $I-V$ characteristic demonstrated.  In that work, the existence of a nonzero $I_c$ was attributed to inhomogeneity in the BSCCO sample but spatially resolved JSTM was not reported.

In this work, we re-examine the problem of JSTM of a $d$-wave superconductor using an $s$-wave tip, accounting for the details of the tunneling process using a Wannier-based framework developed for conventional quasiparticle STM \cite{Choubey2014,Kreisel2015}. We show that while symmetry indeed dictates a vanishing of the {\it average } critical current, there is a large variation of the local critical current within the unit cell. In particular, in BSCCO $I_c$ vanishes above the Cu sites, and displays maximum magnitude near O sites. We further study the effects of Zn-like strong impurity substitution and show that $I_c$ is suppressed in the vicinity of the impurity, owing to the decay of the $d$-wave order parameter\cite{graham2019josephson}. In contrast to the homogeneous case, $I_{c}$ assumes non-zero values above Cu sites near impurity. Moreover, we show that the Fourier transform of the $I_c$-map in a large field of view centered around a weak non-magnetic impurity shows characteristic wavevectors, which are distinct from the energy-integrated quasiparticle interference (QPI) obtained from conventional STM studies \cite{fujita2014simultaneous, Kreisel2015}. These real-space and wavevector-space results constitute our main predictions for signatures of $d$-wave pairing in cuprates for future JSTM experiments measuring $I_c$ using an $s$-wave tip.

Finally, we study tunneling obtained from the dispersing branch of the quasiparticle spectrum in the same {SIS} set-up and show that the tunneling conductance spectrum has coherence peaks at $\pm(\Delta_{s} + \Delta_{t})$, as expected, and a full gap spanning $-\Delta_{t}<\omega<\Delta_{t}$. Moreover, non-magnetic impurity-induced resonance peaks are also shifted by $\Delta_{t}$; however, real-space patterns of tunneling conductance at the shifted energies turn out to be almost identical to the corresponding conventional STM result. 

\section{Model}
\label{sec:Model}
We consider a JSTM set-up with pointlike $s$-wave tip and superconducting BSCCO sample. The $d$-wave superconducting state of the sample is described by the following mean-field Hamiltonian $H$.
\begin{align}
&H = H_{0} + H_\text{SC} + H_\text{imp}, \label{eq:H}\\
&H_{0} = \sum_{i j \sigma} t_{ij}c_{i\sigma}^{\dagger}c_{j\sigma} - \mu_{0}\sum_{i \sigma}c_{i\sigma}^{\dagger}c_{i\sigma}, \label{eq:H0}\\
&H_\text{SC} = \sum_{ij} \Delta_{ij} c_{i\uparrow}^{\dagger}c_{j\downarrow}^{\dagger} + \text{H.c.} \label{eq:H_sc}\\
&H_{\text{imp}} = \sum_{\sigma} V_{\text{imp}}c_{i^{\ast}\sigma}^{\dagger}c_{i^{\ast}\sigma}.\label{eq:H_imp}
\end{align}

Here, $H_{0}$ describes a tight-binding model of the normal state of BSCCO obtained by Wannier downfolding of density functional theory (DFT) results as obtained in Ref. \cite{Choubey2017B}. $c_{i\sigma}^{\dagger}$ creates an electron with spin $\sigma$ at lattice site $\bR_{i}$, $t_{ij}$ is the hopping amplitude, and $\mu_{0}$ is the chemical potential. $H_\text{SC}$ describes $d$-wave superconducting state driven by an attractive nearest-neighbor interaction  leading to pairing mean-fields $\Delta_{ij}$. $H_{\text{imp}}$ models a pointlike substitutional non-magnetic impurity, with potential $V_{\text{imp}}$, located at site $i^{\ast}$. The Hamiltonian in Eq. \ref{eq:H} is solved within the self-consistent Bogoliubov-de Gennes (BdG) framework described in the Supplemental Material \cite{SM}.  

Josephson tunneling current between an $s$-wave tip and the lattice site $\bR$ of the sample is given by $I_{J}(\bR) = I_{c}(\bR) \sin{\Delta\phi}$, where $\Delta\phi$ is the phase difference between superconducting order parameters of the tip and the sample, and $I_{c}$ is the critical Josephson tunneling current, which, to the lowest order in the tunneling amplitude $t_{0}$, can be expressed as \cite{graham2017imaging, Schmalian2018}
\begin{equation}
\label{eq:Ic_lattice}
\begin{aligned}
I_{c}(\bR) = \alpha_{0}\int \frac{d\omega}{2\pi} n_{F}(\omega)\text{Im} \qty[F_{s}^{{{\dagger}}}\qty(\bR, \bR; \omega) F_{t}\qty(\omega)],
 \end{aligned}
\end{equation}
where $\alpha_{0} = 8et_{0}^2/\hbar$ is a constant, $F_{t}$ ($F_{s}$) is the anomalous lattice Green's function of the tip (sample). We can generalize Eq. \ref{eq:Ic_lattice} to continuum space discretized into a fine mesh with several intra-unit cell grid points $\br$ by replacing the anomalous lattice Green's function with its continuum analog,
\begin{equation}
\label{eq:Ic_continuum}
\begin{aligned}
I_{c}(\br) \propto \int \frac{d\omega}{2\pi} n_{F}(\omega)\text{Im} \qty[F_{s}^{{{\dagger}}}\qty(\br, \br; \omega) F_{t}\qty(\omega)],
 \end{aligned}
\end{equation}
where, $F_{s}\qty(\br, \br; \omega)$ is the anomalous local Greens function in continuum space, which can be obtained employing a Wannier basis transformation as \cite{Choubey2014}.
\begin{equation}
\label{eq:Ic_continuum}
\begin{aligned}
F_{s}\qty(\br, \br; \omega) = \sum_{\bR\bRp} F_{s}\qty(\bR, \bRp; \omega)w_{\bR}(\br)w_{\bRp}(\br).
 \end{aligned}
\end{equation}
Here, $w_{\bR}$ is the Wannier function centered at lattice site $\bR$. We have used a DFT-derived Cu-$d_{x^2 - y^2}$ surface Wannier function as obtained in Ref. \cite{Choubey2017B}. To compare JSTM observables with conventional STM results, we compute the sample's continuum local density of states (LDOS) $\rho(\br, \omega)$ using 
\begin{equation}
\label{eq:Ic_continuum}
\begin{aligned}
\rho(\br, \omega) = -\frac{2}{\pi}\text{Im}\qty[G_{s}\qty(\br, \br; \omega)],
 \end{aligned}
\end{equation}
where the continuum Greens function $G_{s}\qty(\br, \br; \omega)$ can be expressed as \cite{Choubey2014,Kreisel2015,Choubey2017B}
\begin{equation}
\label{eq:Ic_continuum}
\begin{aligned}
G_{s}\qty(\br, \br; \omega) = \sum_{\bR\bRp} G_{s}\qty(\bR, \bRp; \omega)w_{\bR}(\br)w_{\bRp}^{\ast}(\br).
 \end{aligned}
\end{equation}
Here, $G_{s}\qty(\bR, \bRp; \omega)$ is the lattice electron Greens function, calculational details of which we provide in the  Supplemental Material \cite{SM}. In addition to $I_{c}(\br)$, the JSTM set-up can also measure quasiparticle tunneling current $I(\br, V)$ between the tip (with density of states $\rho_{t}(\omega)$) and sample as a function of applied bias voltage $V$ and tip position $\br$. $I(\br, V)$ can be expressed as\cite{Tinkham}
\begin{equation}
\label{eq:I_quasiparticle}
\begin{aligned}
I(\br, V) \propto \int_{-\infty}^{\infty}d\omega\rho_{t}(\omega)\rho(\br, \omega + eV)\qty[n_F(\omega) - n_F(\omega + eV)].
 \end{aligned}
\end{equation}

 \begin{figure}
\begin{center}
\includegraphics[width=1\columnwidth]{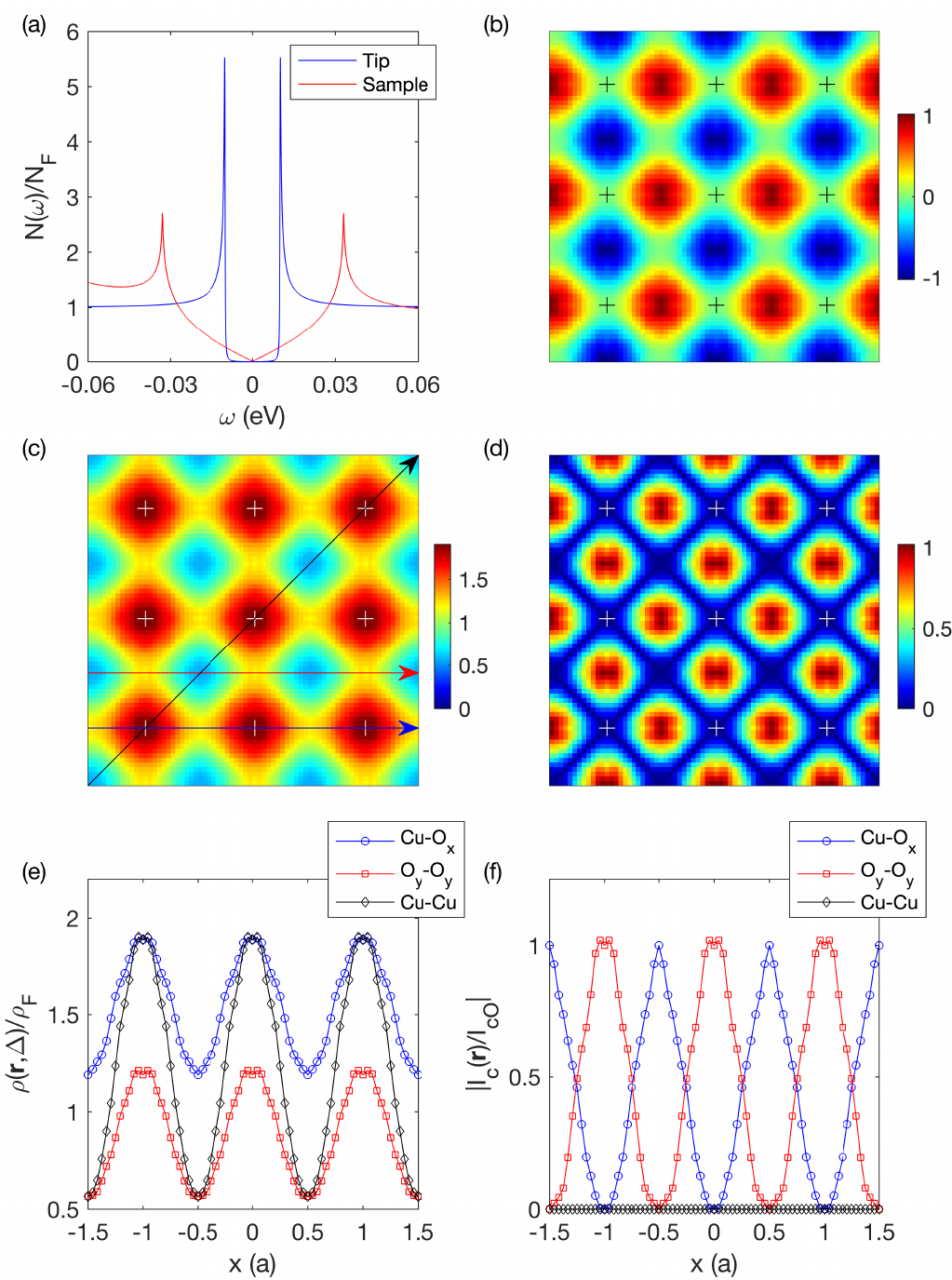}
\caption{Critical Josephson current ($I_{c}(\mathbf{r})$) in the JSTM set-up with an $s$-wave superconducting tip and a clean superconducting BSCCO sample. (a) Lattice DOS spectrum in the tip and the sample, normalized with respect to the normal-state Fermi surface DOS $N_{F}$, showing a spectral gap of $\Delta_{t} = 0.01$ eV and $\Delta_{s} = 0.033$ eV, respectively. $I_{c}(\mathbf{r})$-map (b) and its magnitude (d) at a height $\approx$ 5 {\AA} above the BiO surface, normalized with respect to its magnitude above O sites ($|I_{cO}|$). Cu positions are marked with cross symbols. (c) Continuum LDOS $\rho(\mathbf{r}, \omega)$-map at energy $\omega = \Delta_{s}$, normalized with respect to the Fermi surface LDOS $\rho_{F}$ above Cu positions. Line-cuts of (e) $\rho(\mathbf{r}, \Delta_{s})$- and (f) $I_{c}(\br)$-map (shown in panels (b) and (c), respectively) along Cu-O$_{x}$ (blue), O$_{y}$-O$_{y}$ (red), and Cu-Cu (black) directions, marked by arrows in panel (c). Position $x$ is measured in the unit of in-plane lattice constant $a$.}
\label{fig:IcMap_homogeneous}
\end{center}
\end{figure}

\section{Results}
\label{sec:Results}
We consider a BSCCO sample with 15\% hole-doping and model its superconducting state through an attractive interaction on NN bonds with amplitude $V_{sc} = 0.164$ eV, resulting in a $d$-wave superconducting gap $\Delta_{s}(\bk) = \Delta_{0}\qty(\cos{k_{x}} - \cos{k_{y}})$, which yields coherence peaks at $\pm\Delta_{s} \approx 33$ meV in the density of states plot (blue curve) shown in Fig. \ref{fig:IcMap_homogeneous}(a). We first consider for clarity an $s$-wave tip  with unrealistically large  gap $\Delta_t = 10$ meV (see Fig. \ref{fig:IcMap_homogeneous}(a)). Subsequent results will be shown for an $s$-wave gap of 1.14 meV corresponding to a Nb tip \cite{Bondarenko2015}. We compute $I_{c}(\br)$ in this JSTM set-up, assuming that the tip is located at a height $z \approx 5$ {\AA} above the BiO cleavage plane. Fig. \ref{fig:IcMap_homogeneous}(b) shows map of $I_{c}(\br)$, normalized with its magnitude above the O position. $I_{c}(\br)$ vanishes along the Cu-Cu directions, changes sign under four-fold rotation, and exhibits the largest magnitude in the vicinity of O positions. A lattice-space calculation of Josephson critical current in the same JSTM setup \cite{graham2017imaging} leads to $I_{c}(\bR) = 0$ everywhere since the on-site lattice anomalous Greens function $F_{s}\qty(\bR, \bR; \omega) = \sum_{\bk}F_{s}\qty(\bk, \omega)$ vanishes for a $d$-wave superconductor \cite{Xiang_Wu_2022, Boker2020}, thereby suggesting that a JSTM experiment on a $d$-wave sample with an $s$-wave tip should yield a null result. However, the continuum anomalous Greens function $F_{s}\qty(\br, \br; \omega)$ vanishes only along Cu-Cu directions, where $d_{x^2 - y^2}$ symmetry is exact, and remains nonzero at other positions, leading to non-zero $I_{c}(\br)$. The Supplemental Material \cite{SM} provides an analytical proof for the same. Further, the sign change of $I_{c}(\br)$ under $\pi/2$-rotation is inherited from $\Delta_{s}(\bk)$ and leads to a vanishing average of $I_{c}$ over a unit cell. We emphasize that these results are independent of the particulars of the $s$-wave tip and details of the sample's Wannier function, and thus, can be regarded as signatures of $d_{x^2 - y^2}$ superconducting gap symmetry. Current JSTM technology can only measure $|I_{c}(\br)|${{\cite{Liu2021A}}}, which we show in Fig. \ref{fig:IcMap_homogeneous}(d). It can be easily contrasted with differential tunneling conductance $dI/dV$ measured in a conventional STM experiment which is proportional to the sample's continuum LDOS $\rho(\br, \omega)$. Fig. \ref{fig:IcMap_homogeneous}(c) shows $\rho(\br, \omega = \Delta_{s})$, exhibiting maxima in the vicinity of Cu positions, in sharp contrast to $I_{c}(\br)$ variation. Note that this behavior of $\rho(\br, \omega)$ remains the same for all values of $\omega$. In Fig. \ref{fig:IcMap_homogeneous}(e) and (f), we further show a detailed contrast between $I_{c}(\br)$ and $\rho(\br, \omega)$ by plotting line cuts along Cu-Cu, Cu-O$_{\text{x}}$, and O$_{\text{y}}$-O$_{\text{y}}$ directions.

\begin{figure}
\begin{center}
\includegraphics[width=1\columnwidth]{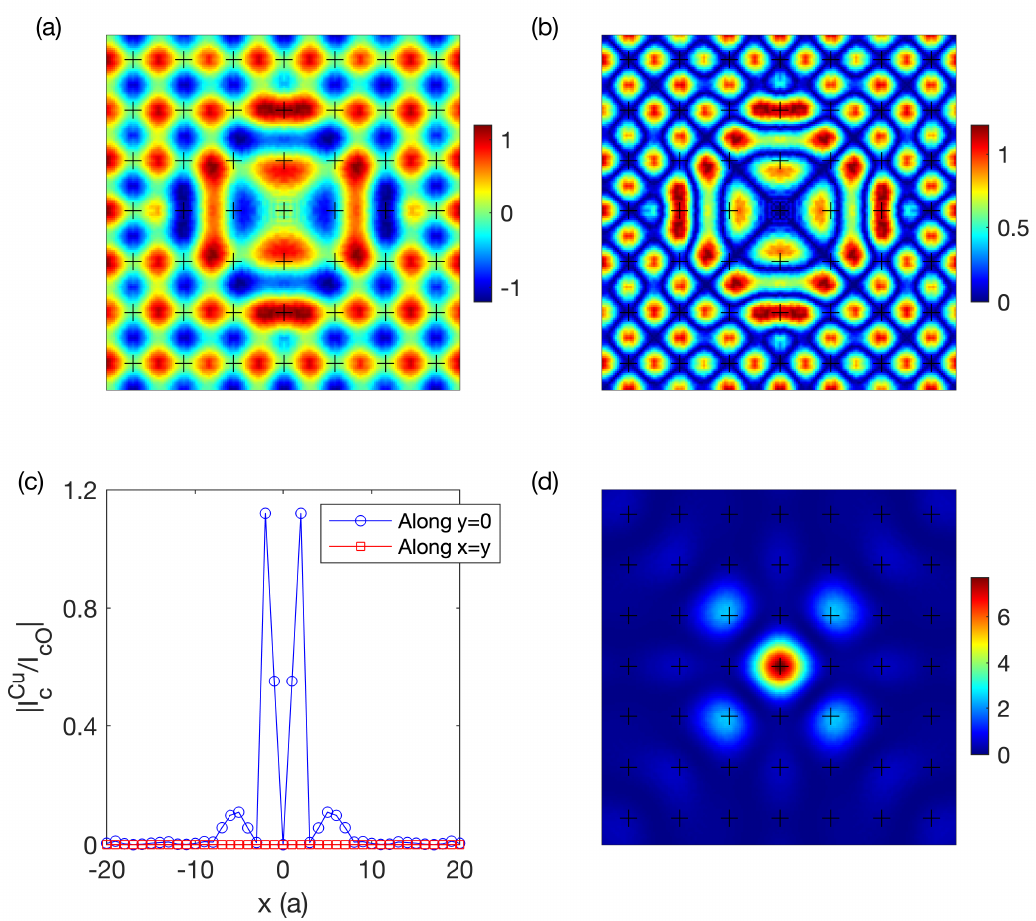}
\caption{$I_{c}(\mathbf{r})$ in the vicinity of a strong, Zn-like, non-magnetic impurity in BSCCO. $I_{c}(\mathbf{r})$-map (a) and its magnitude (b) at a height $\approx$ 5 {\AA} above the BiO surface, normalized with respect to its magnitude above O sites ($|I_{cO}|$) in the homogeneous system. Cu positions are marked with cross symbols, and the impurity, modeled with an on-site potential $V_{\text{imp}} = -5$ eV, is located at the center. (c) $|I_{c}(\mathbf{r})|$ above Cu positions along $x$-axis (blue) and diagonal direction (red). (d) Continuum LDOS ($\rho(\mathbf{r}, \omega)/\rho_{F}$)-map obtained at the impurity-induced resonance energy $\Omega = -4$ meV and within the same spatial region as (a).}
\label{fig:IcMap_Zn}
\end{center}
\end{figure}

Next we study Josephson critical current in the vicinity of a non-magnetic impurity in an JSTM setup with Nb tip. We model a strong, Zn-like, non-magnetic impurity with $V_{\text{imp}} = -5$ eV \cite{Kreisel2015,Choubey2017B} and solve the BdG equations self-consistently. We find impurity-induced resonances at energies $\pm\Omega \approx 4$ meV (see Supplemental Material \cite{SM} for details). Continuum LDOS map $\rho(\br, \omega = -\Omega)$, shown in Fig. \ref{fig:IcMap_Zn}(d), exhibits a maximum { above} the impurity site \cite{Kreisel2015,Choubey2017B} as observed in conventional STM experiments {{\cite{Pan2000}}}. In contrast, $I_{c}(\br)$ above O sites is suppressed  close to the impurity and recovers to its homogeneous value after a few lattice constants as shown in Fig. \ref{fig:IcMap_Zn}(a),(b). The suppression of $I_{c}$ can be attributed to the suppression of the superconducting order parameter near a pair-breaking defect \cite{graham2017imaging,graham2019josephson}. Note that the characteristic behaviors of homogeneous $I_{c}(\br)$ vanishing in diagonal directions and changing sign under $\pi/2$ rotation are preserved even in the presence of the impurity at the origin. More importantly, we find that the presence of impurity leads to non-zero $I_{c}$ over its neighboring Cu positions, in  sharp contrast with the homogeneous case. Fig. \ref{fig:IcMap_Zn}(c) shows that the magnitude of impurity-induced $I_{c}$ can be comparable to its largest value near $O$ sites in the homogeneous case. This effect is a consequence of lattice anomalous Greens functions $F_{s}\qty(\bR, \bR; \omega) \neq 0$ for sites $\bR$ around the impurity which do not lie in the diagonal directions (see Supplemental Material \cite{SM} for more details). Current JSTM experiments {should be able to} measure $|I_{c}(\br)|$ around a Zn impurity in {\BSCCO} and test our predictions (Fig. \ref{fig:IcMap_Zn}(b) and (c)). 

\begin{figure}
\begin{center}
\includegraphics[width=\columnwidth]{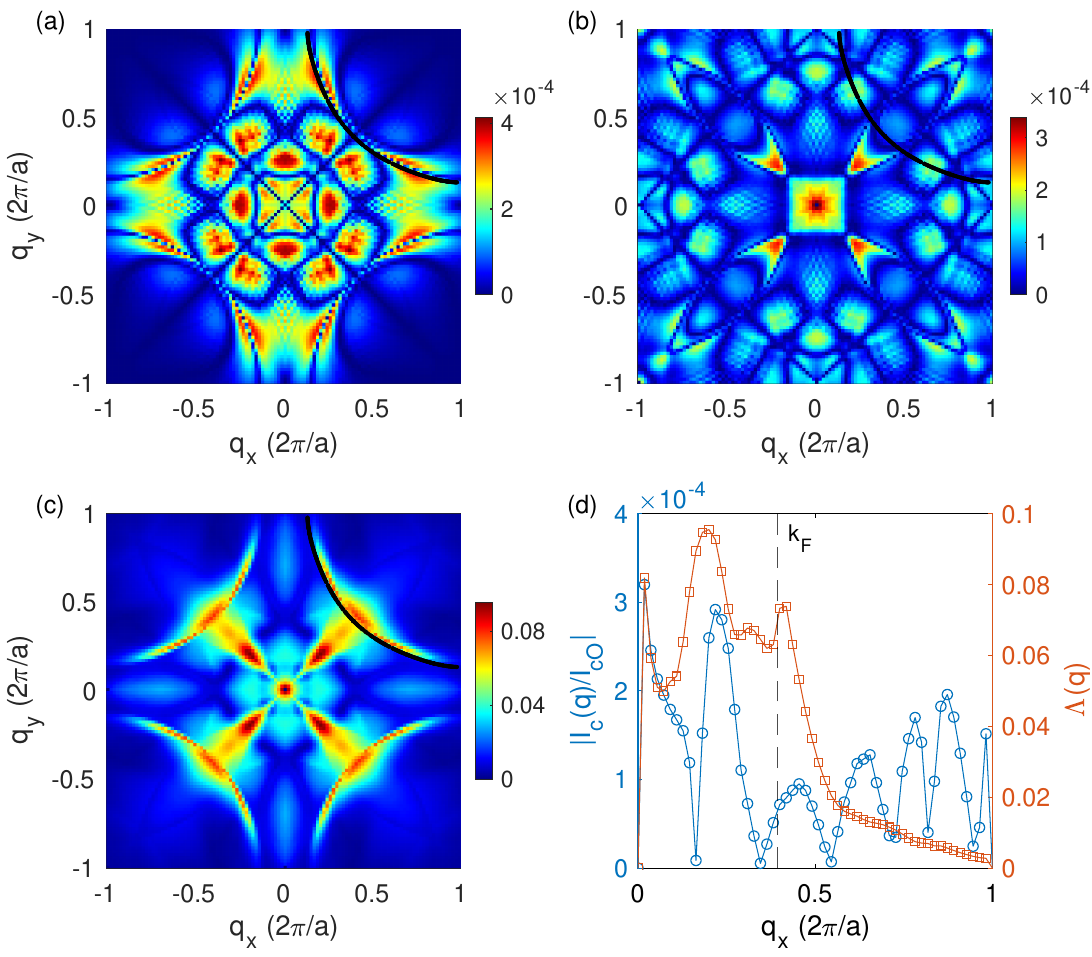}
\caption{Critical Josephson current in wavevector-space. (a) Magnitude spectrum of Fourier transform $I_c(\bq)$ of critical Josephson current maps in presence a weak impurity with $V_{\text{imp}} = 0.3$ eV, used to model native defects in BSCCO. (b) Magnitude spectrum of Fourier transform $|I_c|(\bq)$ of critical current magnitude $|I_c(\br)|$. (c) Energy-integrated QPI $\Lambda(\bq)$-map due to the same impurity. The normal-state Fermi surface in positive quadrant is shown by the black curve in (a)-(c). (d) Line-cuts of panel (b) (blue) and panel (c) (orange) along $q_{x} = q_{y}$-direction. The vertical dashed line in black identifies the Fermi momentum in the same direction.}
\label{fig:IcQ_maps}
\end{center}
\end{figure}

QPI studies have proven instrumental in confirming the $d$-wave symmetry of  the superconducting gap in BSCCO \cite{Fujita2012}. An impurity will give rise to decaying order parameter oscillations similar to the Friedel oscillations of the  LDOS, which can be studied in momentum space to obtain information on important pair scattering processes. We therefore compute  the Fourier transform of the $I_{c}(\br)$ and $|I_{c}(\br)|$ maps in a BSCCO sample with a weak impurity ($V_{\text{imp}} = 0.3$ eV) at the center, which models native defects (Fig. \ref{fig:IcQ_maps}(a) and (b), respectively), and compare it with the energy-integrated QPI ``fingerprint" $\Lambda(\bq)$-maps measured in the conventional STM set-up (Fig. \ref{fig:IcQ_maps}(c)) \cite{Kreisel2015,fujita2014simultaneous}. The Supplemental Material \cite{SM} provides details of the calculation of $\Lambda(\bq)$. The $I_{c}(\bq)$ map shows intense structures around $\bq \approx (0.25, 0)$, $(0.35, 0.25)$ and corresponding symmetry-related wavevectors, and features near $(\pm1, 0)$, $(0, \pm1)$ in regions bounded by the normal-state Fermi surface (Fig. \ref{fig:IcQ_maps}(a)). Most importantly, $I_{c}(\bq)$ vanishes along the nodal directions, in stark contrast with $\Lambda(\bq)$ map (Fig. \ref{fig:IcQ_maps}(c)), which exhibits intense blobs along the same directions. The magnitude spectrum of the Fourier transform of $|I_{c}(\br)|$, which can be measured by present JSTM technology, shows the most intense features around $\bq = (0, 0)$, $(0.25, 0.25)$ and symmetry-related wavevectors (Fig. \ref{fig:IcQ_maps}(b)). Moreover, it has no ``Fermi surface tracing features", unlike $I_{c}(\bq)$ and $\Lambda(\bq)$. These differences are further highlighted by the line-cuts of Fig. \ref{fig:IcQ_maps}(b) and (d) taken along a nodal direction, as shown in Fig. \ref{fig:IcQ_maps}(d). While such QPI features can be understood for a $d$-wave superconductor in the framework of the  ``octet model" \cite{Wang2003}, a theoretical framework explaining the origin of prominent wavevectors in $I_{c}(\bq)$ has not yet been proposed.

\begin{figure}
\begin{center}
\includegraphics[width=1\columnwidth]{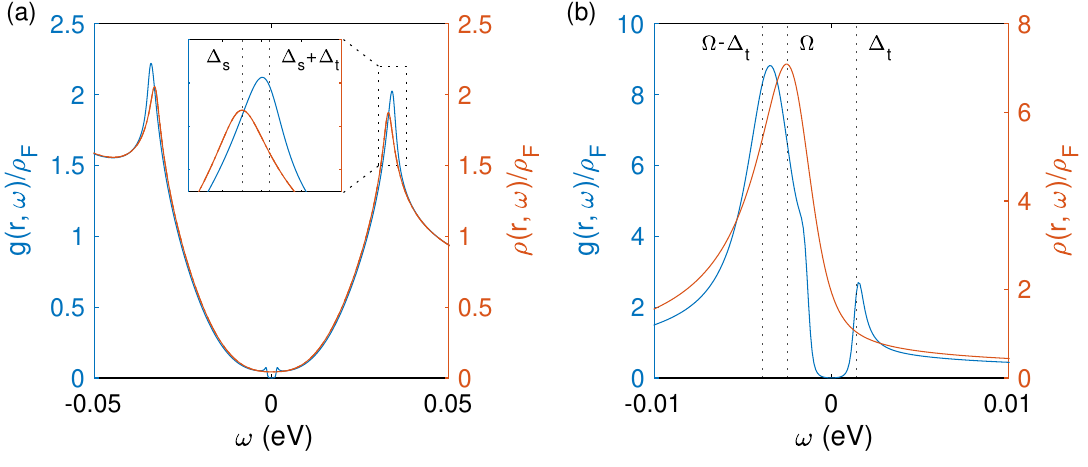}
\caption{Quasiparticle tunneling conductance ($g(\br, \omega)$) in JSTM set-up. (a) $g(\br, \omega)$ (blue) and $\rho(\br, \omega)$ (orange) above Cu positions in the homogeneous state. Both quantities are scaled with the normal-state Fermi energy LDOS $\rho_{F}$. Inset shows zoomed-in view around $\omega = \Delta_{s}$. (b) $g(\br, \omega)$ (blue) and $\rho(\br, \omega)$ (orange) above a Zn-like impurity.}
\label{fig:dIdV_spectra}
\end{center}
\end{figure}

To predict quasiparticle tunneling observables in an JSTM set-up with Nb tip and BSCCO sample, we first compute the sample's continuum LDOS $\rho(\br, \omega)$ using the $T$-matrix formalism (see Supplemental Material \cite{SM} for details, then use Eq. \ref{eq:I_quasiparticle} to calculate tunneling current $I(\br, V)$, and, finally, compute the differential tunneling conductance $g(\br, \omega) = dI/dV|_{\omega = eV}$. The $T$-matrix formalism allows for high energy resolution, necessary to resolve small Nb gap, at a much lower computational cost and without any qualitative difference from direct diagonalization \cite{Choubey2017B}. Fig. \ref{fig:dIdV_spectra}(a) shows $g(\br, \omega)$ above a Cu position in the homogeneous state (normalized by the normal state Fermi level LDOS $\rho_{F}$ at the same position) compared with the continuum LDOS $\rho_{s}(\br, \omega)$. The low-energy spectrum in $g(\br, \omega)$ exhibits a full gap in the range $|\omega|<\Delta_{t}$ and coherence peaks are shifted to $\approx\pm(\Delta_{s}+\Delta_{t})$, which is expected since the SIS tunneling current at low temperatures is simply proportional to the convolution of tip and sample LDOS \cite{hamidian2016detection}. Next, we consider the $g(\br, \omega)$ spectrum in the vicinity of a Zn-like impurity. The $g(\br, \omega)$ spectrum above the impurity site shows a full gap with coherence peak at $\omega = \Delta_{t}$, which is inherited from the superconducting tip (Fig. \ref{fig:dIdV_spectra}(b)). Moreover, the impurity-induced resonance peak gets shifted by an amount $\Delta_{t}$. We find that the real-space pattern at the shifted energies are almost identical to the corresponding continuum LDOS patterns. Furthermore, the analog of the energy-integrated QPI map for quasiparticle tunneling $\Lambda_{g}(\bq)$ too, is very similar to the $\Lambda(\bq)$ map measured in conventional STM experiments, see Supplemental Material \cite{SM} for more details. These predictions can be readily verified in future JSTM experiments.

\section{Conclusions}
\label{sec:Conclusions}
In this work, we have addressed the question whether an $s$-wave superconducting tip can be used to probe a BSCCO sample with $d$-wave gap symmetry in the JSTM set-up. Using first-principles Wannier function, we extended the formalism for computing Josephson critical current in lattice space \cite{graham2017imaging,graham2019josephson} to continuum space, allowing for sub-Angstrom spatial resolution while capturing filtering effects of intervening layers between CuO plane and the STM tip, which have been shown to be crucial for understanding conventional STM results\cite{Kreisel2015}. Contrary to the lattice  results predicting no Josephson signal whatsoever, we find that the continuum critical current remains finite everywhere except along Cu-Cu directions, changes sign under four-fold rotation such that its average over a unit cell vanishes, and attains maximum magnitude around O sites. Further, we showed that, similar to the $d$-wave gap order parameter, $I_{c}(\br)$ is suppressed near a strong non-magnetic impurity. More importantly, the presence of the impurity results in non-zero $I_c$ above neighboring Cu positions. Furthermore, by Fourier transforming $I_{c}(\br)$, we obtained an analog of energy-integrated QPI ($\Lambda(\bq)$) for JSTM set-up and showed that it exhibits characteristic wavevectors which are qualitatively different from that in $\Lambda(\bq)$. Finally, we calculated quasi-particle tunneling conductance in a homogeneous BSCCO sample as well as in the presence of an impurity. We found that the conductance spectra inherits the full gap structure of the tip, and that the sample's coherence peaks as well as impurity-induced resonances are shifted by an amount equal to the tip gap. Moreover, we find that the real-space pattern of the tunneling conductance at the bound state and energy-integrated QPI are very similar to the corresponding conventional STM results. Present JSTM technology can test our predictions of observables related to $|I_{c}(\br)|$ and $g(\br, \omega)$. The formalism presented here is not limited to BSCCO but can be applied to any JSTM set-up with $s$-wave superconducting tip and unconventional superconducting sample. Also, it is straightforward to extend the formalism to include a flake-like tip \cite{graham2019josephson}. Most importantly. we have shown that the orthogonality of the tip and sample's superconducting gap functions does not imply null JSTM results even for a perfectly homogeneous sample.

\section{Acknowledgements}
The authors are grateful for discussions with J.C. Davis,  D. Morr, and J. Paaske. PJH acknowledges support from NSF-DMR-2231821.

\bibliography{references}

\end{document}